\documentclass[aps,prd
,preprint,tightenlines,superscriptaddress,nofootinbib,showpacs]{revtex4}
\usepackage{amssymb,latexsym}
\usepackage{amsmath,amsbsy,bbm}
\usepackage{epsfig,bm}
\usepackage{graphicx,comment}
\DeclareMathOperator{\tr}{tr}

\begin{document}
\def\a{{\alpha}}
\def\b{{\beta}}
\def\d{{\delta}}
\def\D{{\Delta}}
\def\e{{\varepsilon}}
\def\g{{\gamma}}
\def\G{{\Gamma}}
\def\k{{\kappa}}
\def\l{{\lambda}}
\def\L{{\Lambda}}
\def\m{{\mu}}
\def\n{{\nu}}
\def\o{{\omega}}
\def\O{{\Omega}}
\def\S{{\Sigma}}
\def\s{{\sigma}}
\def\th{{\theta}}

\def\ol#1{{\overline{#1}}}

\def\Dslash{D\hskip-0.65em /}
\def\Dtslash{\tilde{D} \hskip-0.65em /}
\def\sumint{{\sum} \hskip-1.1em {\int}}

\def\CPT{{$\chi$PT}}
\def\QCPT{{Q$\chi$PT}}
\def\PQCPT{{PQ$\chi$PT}}
\def\tr{\text{tr}}
\def\str{\text{str}}
\def\diag{\text{diag}}
\def\order{{\mathcal O}}

\def\cC{{\mathcal C}}
\def\cB{{\mathcal B}}
\def\cT{{\mathcal T}}
\def\cQ{{\mathcal Q}}
\def\cL{{\mathcal L}}
\def\cO{{\mathcal O}}
\def\cA{{\mathcal A}}
\def\cQ{{\mathcal Q}}
\def\cR{{\mathcal R}}
\def\cH{{\mathcal H}}
\def\cW{{\mathcal W}}
\def\cM{{\mathcal M}}
\def\cD{{\mathcal D}}
\def\cN{{\mathcal N}}
\def\cP{{\mathcal P}}
\def\cK{{\mathcal K}}
\def\Qt{{\tilde{Q}}}
\def\Dt{{\tilde{D}}}
\def\St{{\tilde{\Sigma}}}
\def\cBt{{\tilde{\mathcal{B}}}}
\def\cDt{{\tilde{\mathcal{D}}}}
\def\cTt{{\tilde{\mathcal{T}}}}
\def\cMt{{\tilde{\mathcal{M}}}}
\def\At{{\tilde{A}}}
\def\at{{\tilde{\alpha}}}
\def\cNt{{\tilde{\mathcal{N}}}}
\def\cOt{{\tilde{\mathcal{O}}}}
\def\cPt{{\tilde{\mathcal{P}}}}
\def\cI{{\mathcal{I}}}
\def\cJ{{\mathcal{J}}}

\def\eqref#1{{(\ref{#1})}}

\title{Current Renormalization in Finite Volume}

\author{Jie Hu}
\email[]{hujie@phy.duke.edu}
\affiliation{Department of Physics, Duke University, Box 90305, Durham, NC 27708-0305, USA}
\author{Fu-Jiun Jiang}
\email[]{fjjiang@itp.unibe.ch}
\affiliation{Institute for Theoretical Physics, Bern University, Sidlerstrasse 5, CH-3012 Bern, Switzerland}
\author{Brian C.~Tiburzi}
\email[]{bctiburz@phy.duke.edu}
\affiliation{Department of Physics, Duke University, Box 90305, Durham, NC 27708-0305, USA}

\date{\today}


\begin{abstract}
For finite volume field theories 
with discrete translational invariance, 
conserved currents can be additively 
renormalized by infrared effects. 
We demonstrate this for pions using chiral perturbation theory 
coupled to electromagnetism in a periodic box. 
Gauge invariant single particle effective theories 
are constructed to explain these results.
In such theories, current renormalization arises 
from operators involving the zero mode of the gauge field.
No contradictions with Ward identities, or low-energy theorems
are encountered. 
\end{abstract}
\maketitle

\section{Introduction}

Since Wilson's pioneering work~\cite{Wilson:1974sk}, 
there has been considerable
activity to solve field theories non-perturbatively by numerical simulation  
on Euclidean spacetime lattices. Today lattice gauge theory is a mature field,
and current state-of-the-art lattice QCD calculations 
are beginning to confront the challenges provided 
by the hadron spectrum. For an overview of lattice methods, see~\cite{DeGrand:2006aa}.
One aspect to these numerical simulations is the finite-size scaling 
of observables. The finite spacetime volume employed on the lattice is 
a source of systematic error in the numerical determination of observables.
Thus the study of field theories in finite volume, while a theoretical
curiosity, is also of practical utility.

Recent work~%
\cite{Detmold:2006vu} 
suggests that electromagnetically
gauge invariant amplitudes at finite volume may differ from their
infinite volume form. Specifically investigated was
the finite-size scaling of nucleon electromagnetic and spin 
polarizabilities that arise in nucleon Compton scattering (see, e.g., 
\cite{Hyde-Wright:2004gh,Schumacher:2005an}).
A goal in 
\cite{Detmold:2006vu} 
was to address systematic errors in the extraction of polarizabilities
from classical background field methods employed in lattice simulations~%
\cite{Fucito:1982ff,Martinelli:1982cb,Bernard:1982yu,Fiebig:1988en,Aoki:1989rx,Aoki:1990ix,Burkardt:1996vb,Christensen:2004ca,Detmold:2004kw,Lee:2005dq,Shintani:2006xr}.
An analysis of the finite volume behavior of nucleon polarizabilities was presented,
as was an oddity relating to the zero-frequency scattering amplitude.
In infinite volume, the zero-frequency Compton amplitude is
fixed by gauge invariance to be proportional to the total charge squared.
Finite volume modifications, however, were found for nucleon 
Compton scattering at zero frequencies~\cite{Detmold:2006vu}.
These results suggest a finite volume renormalization of the basic interaction
between the photon and the hadron's charge. 
In this work, we show that gauge invariance in finite volume
allows for such modifications to zero-frequency photon couplings. 
In essence, conserved currents are not protected from additive 
renormalization as they are in infinite volume. 
For definiteness, we focus on the chiral dynamics of pions
coupled to photons~\cite{Gasser:1983yg}, but could just as well choose 
any interacting field theory coupled to gauge fields.\footnote{%
An instructive alternate example is the QED electron. 
Straightforward evaluation shows that the 
electron vertex function at zero frequency is modified by volume effects. 
This modification, however, is infrared divergent
and we have chosen to avoid such difficulties by 
using a theory that is infrared finite.
}

Our presentation is organized as follows. 
First in Sec.~\ref{s:pions}, we analyze
the electromagnetic interactions of pions 
in finite volume. We demonstrate the
infrared running of electromagnetic 
current matrix elements by explicit one-loop calculations
in chiral perturbation theory (\CPT). 
In Sec.~\ref{s:tori}, consequences of 
gauge invariance on a torus are detailed. 
Gauge invariant zero-mode interactions 
allow for infrared renormalization of
electromagnetic couplings. 
We write down 
gauge invariant, zero-frequency 
effective field theories for pions
that reproduce our one-loop finite volume \CPT\ results.
Understanding such volume effects is  
necessary in practice for the extraction 
of infinite volume 
physics from lattice QCD simulations.
We show how our results are consistent with 
Ward identities and low-energy theorems
in Sec.~\ref{s:identity}.
A conclusion in Sec.~\ref{s:summy}
summarizes our findings, while
a glossary of finite volume functions
is provided in the Appendix.

\section{Pions in Finite Volume} \label{s:pions}

The chiral Lagrangian is written in terms of a coset
field $\Sigma = \exp ( 2 i \Phi / f)$ which parametrizes
the Goldstone manifold arising from spontaneous
chiral symmetry breaking: $SU(2)_L \otimes SU(2)_R \to SU(2)_V$. 
The pions are contained in the matrix $\Phi$, explicitly as
\begin{equation} \label{eq:pions}
\Phi 
= 
\begin{pmatrix}
\frac{1}{\sqrt{2}} \pi^0  &                    \pi^+ \\
\pi^-                     & - \frac{1}{\sqrt{2}} \pi^0
\end{pmatrix}
.\end{equation}
In our conventions, the dimensionful parameter 
$f = 132 \, \texttt{MeV}$. 
The chiral Lagrangian provides an effective theory
of low-energy QCD. At leading-order 
in an expansion in momentum, $p^2$, and quark mass, $m_q$, 
there are two terms in this Lagrangian
\begin{equation} \label{eq:L}
\cL 
= 
\frac{f^2}{8} 
\tr \left( D_\mu \Sigma D^\mu \Sigma^\dagger \right) 
+
\frac{f^2}{4} \lambda \,
\tr \left( \Sigma m_q + \Sigma^\dagger m_q \right)
,\end{equation}
where $m_q$ is the quark mass matrix, $m_q = \diag (m_u, m_d)$. 
We shall work exclusively in the isospin limit, $m_u = m_d \equiv m$. 
The kinetic term of the chiral Lagrangian includes a
$U(1)$ gauge covariant derivative that couples pions
to photons,
$D_\mu \Sigma = \partial_\mu \Sigma + i e A_\mu \left[ Q, \Sigma \right]$,
where the quark electric charge matrix, $Q$, is given by $Q = \diag (2/3, -1/3)$.

Expanding the Lagrangian in Eq.~\eqref{eq:L} to tree level, 
one sees that the pions are correctly normalized and 
their mass, $m_\pi$, is given by 
$m_\pi^2 = 2 \lambda m$. 
The couplings of pions to zero-momentum photons 
at tree level can be read off from Eq.~\eqref{eq:L}, 
from which we find their canonical charges. 
We now investigate whether loop corrections 
in a finite spatial volume modify these couplings.

\subsection{Charged pion current}

\begin{figure}[tb]
\bigskip
  \includegraphics[width=0.13\textwidth]{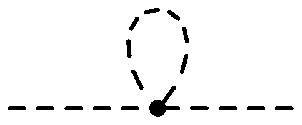}%
$\quad$
$\quad$
$\quad$
$\quad$
  \includegraphics[width=0.13\textwidth]{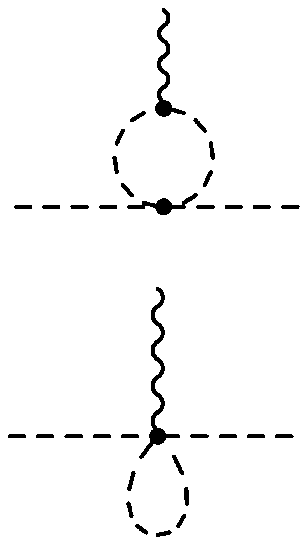}%
  \caption{
One-loop graphs required to evaluate the pion 
electromagnetic current.
On the left appears the wavefunction correction; 
while, on the right, diagrams contributing to the 
pion form factor.
Vertices shown are generated from the leading-order \CPT\ Lagrangian, Eq.~\eqref{eq:L}. 
  }
  \label{f:Charge}
\end{figure}

To consider the one-loop corrections 
to the electromagnetic current of charged pions, 
we accordingly expand the \CPT\ Lagrangian in Eq.~\eqref{eq:L} 
to second order to generate vertices for one-loop graphs. 
Furthermore local terms at higher-order can then contribute at 
tree-level, but these are absent for zero-frequency photons.
Thus we need to determine only the diagrams depicted in Fig.~\ref{f:Charge}.

In the limit of zero frequency and infinite volume, 
the current matrix element
between charged pion states is required by gauge and Lorentz invariance 
to be  
\begin{equation} \label{eq:chargevertex}
\langle \pi^\pm (P) | J_\mu | \pi^\pm (P) \rangle =  \pm 2 e P_\mu 
,\end{equation}
where the overall sign reflects the charge of the pion. 
It is a straightforward exercise to verify the above form 
at one-loop order in infinite volume. 
At an intermediate step, we reach the result
\begin{equation} \label{eq:chargestep}
\langle \pi^\pm (P) | J_\mu | \pi^\pm (P) \rangle 
=  
\pm 2 e  
\left\{
P_\mu
- \frac{2 i}{f^2}
\int \frac{d^4 q}{(2 \pi)^4} 
\frac{(q^2 - m_\pi^2) P_\mu - 2  (q \cdot P) q_\mu  }{[q^2 - m_\pi^2]^2}
\right\}
,\end{equation}
from which we see the wavefunction correction to the tree-level 
vertex is exactly canceled by the loop contributions 
to the form factor at zero frequency~\cite{Gasser:1984ux}.
In this way, the matter fields do not contribute to the 
running of the coupling and Eq.~\eqref{eq:chargevertex} is preserved.

In finite volume, we repeat the calculation of the 
pion current to one-loop order. We consider each of the three spatial
directions of finite length $L$, and the quark fields subjected
to boundary conditions that maintain discrete translational invariance. 
For definiteness, we assume periodic boundary conditions.\footnote{%
Similar results for anti-periodic boundary conditions, for example, 
can be derived easily using a modified momentum quantization condition
for the quark fields. As pions remain periodic, the expressions we derive 
also hold for anti-periodic quarks.
} 
As the pions are point-like objects
in the effective theory, they satisfy the same boundary conditions
as the point-like interpolating field $\Phi(x) \sim \ol q(x) \gamma_5 q(x)$.
Pions are hence also periodic with quantized spatial momentum modes of the form 
\begin{equation}
\bm{q} = \frac{2 \pi}{L} \bm{n}
,\end{equation}
where $\bm{n}$ is a triplet of integers. 
To keep matters simple, we keep the temporal extent infinite as is commonly 
done to determine finite-size effects
for lattice QCD observables.\footnote{%
We implicitly choose $m_\pi L \gtrsim 1$
so that pion zero modes do not become strongly coupled~\cite{Gasser:1986vb,Gasser:1987ah}.
With this assumption, 
the ordinary \CPT\ power counting in infinite volume
can be carried over to finite volume~\cite{Gasser:1987zq}. 
}
To evaluate the pion current, 
we use the finite volume theory defined by Eq.~\eqref{eq:L}.
The loop diagrams shown in Fig.~\ref{f:Charge} are again generated. 
The only difference compared to infinite volume 
is that the spatial momenta of real and virtual states are quantized.
The finite and infinite volume theories share exactly the same 
ultraviolet divergences, so we can calculate the finite volume
effect by matching the two theories in the infrared. 
For an observable $X$ calculated in both finite, $X(L)$, and 
infinite, $X(\infty)$, volumes, we have
\begin{equation} \label{eq:cartoon}
X(L) = X(\infty) + \Delta X(L)
,\end{equation}
where the matching term, $\Delta X(L)$, is free from ultraviolet 
divergences and gives the finite volume effect.

\begin{figure}[tb]
	\bigskip		
  \includegraphics[width=0.38\textwidth]{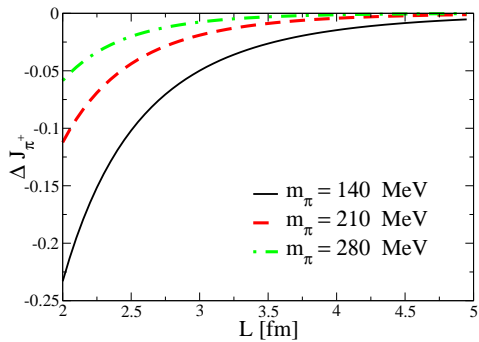}%
  \caption{Finite volume screening of the pion current.
  The relative difference in pion current $\Delta J_{\pi^+}$ is plotted 
  as a function of the box size $L$, for a few values of the
  pion mass.  
  }
  \label{f:charge}
\end{figure}

Returning to Eq.~\eqref{eq:chargestep}, we can carry out the finite volume matching, 
Eq.~\eqref{eq:cartoon}, for the pion current. We find
\begin{equation} \label{eq:charged}
\langle \pi^\pm (P) | J_\mu | \pi^\pm (P) \rangle =  \pm 2 e P_\mu 
\left\{
1 
- 
\frac{\delta_{\mu j}}{3 f^2} 
\left[
2 \, I_{1/2}(m^2_\pi, L) + m_\pi^2 I_{3/2}(m^2_\pi, L)
\right]
\right\}
,\end{equation}
where $I_{\b}(m^2, L)$ is  
defined in the Appendix.
Results are consistent with charge conjugation invariance 
and the current is only modified in the spatially 
finite directions. Specifically the virtual pion cloud
in finite volume screens the current
of the infinite volume pion. 
In Fig.~\ref{f:charge}, we plot the finite volume modification to the 
pion current. 
Here the relative difference in the current matrix element 
at finite volume versus infinite volume, $\Delta J_{\pi^+}$,
given by
\begin{equation}
\Delta J_{\pi^+} 
= 
\frac{
\langle \pi^+ (P) | \hat{\bm{e}} \cdot \bm{J} | \pi^+ (P) \rangle_L 
-  
\langle \pi^+ (P) | \hat{\bm{e}} \cdot \bm{J} | \pi^+ (P) \rangle_\infty
}
{\langle \pi^+ (P) | \hat{\bm{e}} \cdot \bm{J} | \pi^+ (P) \rangle_\infty}
,\end{equation}
is plotted as a function of the length $L$ of the spatial dimension. 
We have used a unit vector $\hat{\bm{e}}$ to project onto
the spatial part of the current. Accordingly the 
pion cannot be at rest, $\bm{P} \neq \bm{0}$. Subscripts on
matrix elements denote the box size, with infinity corresponding
to infinite volume.
The finite volume effect is exponentially suppressed in 
asymptotic ($m_\pi L \gg 1$) volumes. 
Consequently taking the infrared cutoff, $1/L$, to zero, 
the additive current renormalization vanishes
and infinite volume limit is maintained.

\subsection{Neutral pion current matrix elements}

Charge conjugation invariance demands the identical vanishing of  
single current matrix elements between neutral pion states.
Indeed whether the calculation of the neutral pion
current is carried out in infinite or finite volume, 
we find zero for the matrix element. The $SU(2)$ flavor structure
of the form factor diagrams shown in Fig.~\ref{f:Charge}
ensures this vanishing and consistency with charge conjugation.

Neutral pion matrix elements of an even number of electromagnetic 
currents, however, are not restricted to vanish by charge conjugation invariance. 
Indeed, it is well known that the neutral pion has electric and 
magnetic polarizabilities that can be predicted at one-loop
order in \CPT\ solely in terms of $f$ and $m_\pi$~\cite{Holstein:1990qy}. 
Such polarizabilities arise at second order in the low-frequency expansion
of the matrix element of two currents (the so-called Compton
scattering tensor). 
The Compton tensor also has a term at zeroth order in the photon
frequencies
\begin{equation} \label{eq:comptontensor}
T_{\mu \nu} (\omega = \omega' = 0) 
= 
2 (Q e)^2 g_{\mu \nu} 
,\end{equation}
which is sensitive only to the longest ranged electromagnetic interaction. 
This term in the Compton tensor, when combined with relevant phase
space factors, 
yields the classical
Thomson scattering cross section, $\sigma = 8 \pi (Q e)^4 / 3 m^2_\pi$. 
For the neutral pion, the total charge is zero
and the longest ranged interaction vanishes.

\begin{figure}[tb]
\bigskip
  \includegraphics[width=0.4\textwidth]{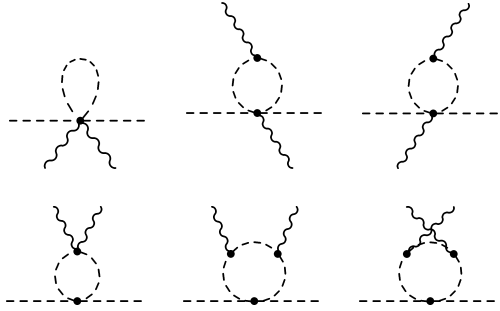}%
  \caption{
One-loop contributions to neutral pion 
Compton scattering in \CPT.
Vertices shown are generated from the 
leading-order Lagrangian. 
  }
  \label{f:Compton}
\end{figure}
Using the \CPT\ Lagrangian defined in Eq.~\eqref{eq:L}, 
we can determine the Compton amplitude for pions. 
We restrict our attention to the zero-frequency amplitude. 
Due to charge neutrality, there are no tree-level
couplings to the neutral pion.
At one-loop order, evaluation of the diagrams 
shown in Fig.~\ref{f:Compton} is required to determine the 
Compton amplitude.
At an intermediate step in the calculation, 
contributions from all six diagrams can be simplified to
\begin{equation} \label{eq:compton}
T_{\mu \nu} (\omega = \omega' = 0) 
=
\frac{4 i e^2 m_\pi^2}{f^2}
\int \frac{d^4 q}{(2 \pi)^4}
\frac{(q^2 - m_\pi^2) g_{\mu \nu} - 4 q_\mu q_\nu}{[q^2 - m_\pi^2]^3}
,\end{equation}
which vanishes. Hence in infinite volume, 
a delicate cancellation between all diagrams maintains the vanishing of 
the Compton amplitude at zero frequency~\cite{Bijnens:1987dc,Donoghue:1988ee}. 
On the other hand, the same is not true in finite volume. 
Carrying out the one-loop matching between finite and infinite volume theories, Eq.~\eqref{eq:cartoon}, 
for the Compton amplitude in Eq.~\eqref{eq:compton}, we find
\begin{equation} \label{eq:neutral}
T_{\mu \nu} (\omega = \omega' = 0) 
= 
e^2 
(\delta_{\mu 0} \delta_{\nu 0} - g_{\mu \nu})
\frac{m_\pi^4}{f^2} \, I_{5/2}(m^2_\pi, L)
.\end{equation}
Thus when one considers the purely spatial 
components of the Compton tensor, the 
neutral pion has an effective charge-squared, 
\emph{cf}.~Eq.~\eqref{eq:comptontensor}. 
\begin{figure}[tb]
	\bigskip		
  \includegraphics[width=0.38\textwidth]{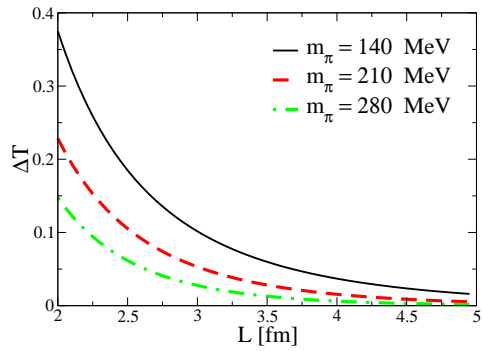}%
  \caption{Zero frequency Compton amplitude for the neutral pion.
  The finite volume amplitude $\Delta T$ is plotted 
  as a function of the box size $L$, for a few values of the
  pion mass.  In infinite volume, this amplitude is 
identically zero.
  }
  \label{f:chargesquared}
\end{figure}
In transverse gauge, the above expression gives 
the amplitude to scatter zero-frequency photons
off the neutral pion. There is a non-vanishing 
contribution to this scattering amplitude
when the pion is confined to a periodic box
with size on the order of the pion Compton 
wavelength. When the box size becomes large
compared to this scale, 
the amplitude is exponentially suppressed
and infinite volume results are recovered. 
We demonstrate this in Fig.~\ref{f:chargesquared}, 
where we plot the finite volume amplitude $\Delta T$
defined by
\begin{equation}
\Delta T = - \frac{1}{3 e^2}  g^{\mu \nu} T_{\mu \nu} (\omega = \omega' = 0) 
.\end{equation}

\section{Gauge invariance on a torus} \label{s:tori}

To explain our above results, we investigate electromagnetism in finite volume.
The analogous finite temperature case is well known and described, e.g., 
in~\cite{Zinn-Justin:1996cy}. 
Because our applications
are with classical background fields, or equivalently current operator
insertion methods in lattice field theory, there are no quantum corrections
to the photon field itself.\footnote{%
Dynamical photons in QED 
cause additional complications 
as the vector current is renormalized in infinite volume~\cite{Collins:2005nj}.
With classical background fields, penguin graphs are absent
and such renormalization does not occur.
}

\subsection{Spatial Torus}
Let us consider a classical electromagnetic field defined on a finite spatial torus
with infinite time extent.
On the gauge field $A_\mu(\bm{x},t)$, we impose periodic boundary conditions 
and expand $A_\mu(\bm{x},t)$ in Fourier modes 
\begin{equation}
A_\mu(\bm{x},t) 
= 
\sum_{\bm{n}} \tilde{A}_\mu (\bm{n},t)
\, 
e^{2 \pi i \bm{n} \cdot \bm{x} / L}
,\end{equation}
where $\bm{n} = (n_x, n_y, n_z) \in \mathbb{Z}^3$. 
It is convenient to separate out the zero-mode contribution, 
so we write
\begin{equation}
A_\mu(\bm{x},t) 
= 
\cA_\mu(t)
+
\ol A_\mu(\bm{x},t)
,\end{equation}
where the zero mode $\cA_\mu(t) \equiv \tilde{A}_\mu(\bm{0},t)$.

Under a gauge transformation, the photon field transforms in the familiar way, 
$A_\mu(\bm{x},t) \longrightarrow A_\mu(\bm{x},t) + \partial_\mu \alpha(\bm{x},t)$, 
and observables are invariant. 
Requiring the gauge transformed field to be single valued 
mandates that $\partial_\mu \alpha(\bm{x},t)$ is periodic.
Thus we can decompose the gauge function $\alpha(\bm{x},t)$ into the sum of two terms,
$\alpha(\bm{x},t) = \alpha_0(\bm{x},t) + \ol \alpha(\bm{x},t)$, 
where
\begin{equation} \label{eq:zeromode}
\alpha_0(\bm{x},t) = \alpha_0(t) + \bm{\alpha} \cdot \bm{x}
,\end{equation}
and
\begin{equation} \label{eq:nonzero}
\ol \alpha(\bm{x},t) 
= 
\sum_{\bm{m} \neq \bm{0}} 
\ol \at (\bm{m},t) \, 
e^{2 \pi i \bm{m} \cdot \bm{x} / L}
.\end{equation}
Here we have dropped all overall irrelevant constants,
and the vector $\bm{\alpha}$ is a constant vector. 
Using this decomposition for the gauge function, 
the gauge field transforms as
\begin{equation} \label{eq:transf}
\begin{cases}
\cA_\mu(t) 
& \longrightarrow  
\cA_\mu(t) 
+ 
\partial_\mu \alpha_0(\bm{x},t) 
\\
\ol A_\mu(\bm{x},t) 
& \longrightarrow 
\ol A_\mu(\bm{x},t)  
+
\partial_\mu \ol \alpha(\bm{x},t)
\end{cases}
.\end{equation}
In particular, the photon zero-mode 
transforms as 
\begin{equation}
\cA_\mu(t)
\longrightarrow
\cA_\mu(t)
+ 
\begin{cases}
\partial_0 \alpha_0(t) , & \mu = 0 \\
\alpha_i , & \mu = i
\end{cases}
.\end{equation}
The time-component of the zero mode is absent from the field strength tensor.
The remaining three components of the zero mode field 
are translated by a constant under the gauge transformation. 
In the gauge invariant free theory, each spatial component of the zero mode
is thus a massless one-dimensional scalar.

\subsection{Coupling to Matter}

For a generic matter field $\varphi(\bm{x},t)$
of unit charge, the effects of a gauge 
transformation show up as a local phase factor
\begin{eqnarray} \label{eq:gaugetransf}
\varphi(\bm{x},t) 
\longrightarrow 
\varphi'(\bm{x},t) 
&=&
e^{- i \alpha(\bm{x},t)} \varphi(\bm{x},t)
.\end{eqnarray}
Now we assume that the matter field
$\varphi(\bm{x},t)$ is subject to periodic boundary conditions.
We again split the gauge function into zero mode and non-zero mode pieces,
$\alpha(\bm{x},t) = \alpha_0(\bm{x},t) + \ol \alpha(\bm{x},t)$.
With the form given in Eq.~\eqref{eq:nonzero}, we see that the non-zero modes 
will maintain the periodicity of the matter field under the transformation in 
Eq.~\eqref{eq:gaugetransf}.
The same is not in general true of the zero modes given the form of
$\a_0(\bm{x},t)$ in Eq.~\eqref{eq:zeromode}.
If the gauge transformed matter field is to remain periodic under translations
by $L$, 
then we must have
the quantization condition
\begin{equation} \label{eq:key}
\bm{\alpha} = \frac{2 \pi}{L} \bm{n}
,\end{equation}
on the spatial zero mode part of the gauge function.
This quantization condition reduces
the continuous translational invariance of the spatial zero modes
to discrete translations. 
As gauge transformations are now less general compared to infinite volume,
more gauge invariant operators can be built.

Imagine that we start with some microscopic theory
with electromagnetic interactions. Take the scalar field 
$\varphi(\bm{x},t)$ as some composite 
low-energy degree of freedom of this theory. 
Further we assume that the energies
of interest are ultra-low in the sense that any interactions of 
$\varphi(\bm{x},t)$ with itself or other fields have been integrated out. 
In the absence of electromagnetism, e.g., we have a simple single particle
effective theory\footnote{%
We have written only $SO(4)$ symmetric terms in Eq.~\eqref{eq:trivial}.
Strictly speaking volume corrections will reduce the dispersion relation 
down to only cubic symmetry. For pions in finite volume the first $SO(4)$ breaking effects occur 
at two-loop order in the chiral expansion.  
}
\begin{equation} \label{eq:trivial}
\cL = |\partial_\mu \varphi|^2 - m(L)^2 |\varphi|^2 
,\end{equation}
where $m(L)$ is a running mass that depends on the infrared cutoff $1/L$
(and parametrically depends on the other couplings, masses, etc.~that have been integrated out of the theory). 
Running the cutoff to zero completes the infrared sector of the theory
and produces infinite volume physics.

Now we include electromagnetism in this single particle effective theory by 
adding all possible gauge invariant operators. 
The minimal coupling prescription, 
$\partial_\mu \to D_\mu = \partial_\mu + i  A_\mu$,
renders the kinetic term of Eq.~\eqref{eq:trivial} gauge invariant. 
Because we imagine 
$\varphi$ 
is a composite
particle, there can be non-minimal couplings that respect gauge invariance, e.g., the 
$\bm{E}^2$
and 
$\bm{B}^2$ 
terms for the particle's polarizabilities. 
Further terms in this ultra low-energy theory are allowed, however, because 
$SO(4)$ 
is not respected, and the gauge invariance of the zero mode has a special nature. 
As we will show, these further terms are responsible 
for current renormalization. 
To simplify the discussion, we will restrict
ourselves to the effective theory operators 
for zero frequency photons.

Using gauge symmetry, we can write down 
the general form of the ultra low-energy effective theory for 
a single $\varphi$ field coupled to zero frequency photons. 
We choose to construct this theory using Wilson lines.
By cycling once over the $i$-th compact dimension, we can form 
gauge invariant Wilson lines $\cW_i$ of the form,
\begin{equation}
\cW_i = \exp \left ( i \oint dx_i \, A_i \right) 
.\end{equation}
Notice that there is no sum over repeated indices in this definition.
Due to the periodicity of the gauge field in the $i$-th direction, 
the loop integral 
\begin{equation}
\int_0^L dx_i \, A_i(\bm{x},t) 
= 
L \mathcal{A}_i(t) 
+ 
L \sum_{ n_{j \neq i}, \,  n_i = 0}
\tilde{A}_i(\bm{n},t) 
e^{2 \pi i \bm{n} \cdot \bm{x} / L},
\end{equation}
produces just the 
$n_i = 0$ 
modes of the gauge field.
Indeed the gauge transformation of the zero and non-zero modes, 
Eq.~\eqref{eq:transf},
demonstrates that the Wilson line 
$\cW_i$ 
is gauge invariant.
For our purpose, we wish to isolate completely the gauge field 
zero-mode and accordingly form modified Wilson lines 
$W_i$ 
given by
\begin{equation}
W_i = \cP_{\bm{0}} \cW_i \cP_{\bm{0}}^\dagger
,\end{equation}
where $\cP_{\bm{0}}$ is an operator that projects onto 
the zero-mode of the gauge field. A practical way
to implement the action of $\cP_{\bm{0}}$
is to change the loop integration
\begin{equation}
\oint dx_i \longrightarrow \frac{1}{L^2} \oint \prod_{j=1}^3 dx_j \equiv \frac{1}{L^2} \oint d\bm{x}
\end{equation}
so that
\begin{equation}
W_i = \exp \left ( i \oint d\bm{x} \, A_i / L^2 \right) 
.\end{equation}
Furthermore it is useful to define 
Hermitian combinations of modified 
Wilson lines that transform simply under 
parity and charge conjugation,
\begin{eqnarray}
W^{(+)}_i &=& \frac{1}{2} \left( W_i + W_i^\dagger \right) 
\\
W^{(-)}_i &=& \frac{1}{2 i} \left( W_i - W_i^\dagger \right) 
.\end{eqnarray}
Notice that because
$W_i^{(+)} = \sqrt{ 1 - [W_i^{(-)}]^2}$, 
any operator involving $W_i^{(+)}$ can be traded
in for a tower of operators involving $W_i^{(-)}$.
Hence we can build our theory solely in terms of $W_i^{(-)}$
operators.

In addition to gauge, $C$, $P$, and $T$ invariance, 
the theory on a torus has $S_4$ cubic invariance. 
Writing down operators consistent with these symmetries, 
we arrive at the following ultra low-energy 
effective Lagrangian for a single $\varphi$ field
\begin{eqnarray} \label{eq:fullmess}
\cL 
&=& 
|D_\mu \varphi|^2 - m(L)^2  |\varphi|^2 
+
\cQ_1 (L) \, 
\bm{W}^{(-)} \cdot \bm{J}
+
\cQ_2  (L) \, 
\left(
\bm{W}^{(-)}
\cdot
\bm{W}^{(-)}
\right)
|\varphi|^2
\notag \\
&&+
\cQ_3 (L) \,
\left(
\bm{W}^{(-)} \cdot \bm{D} \varphi^*
\right)
\left(
\bm{W}^{(-)} \cdot \bm{D} \varphi
\right)
+ 
\cQ_4 (L) \,
\sum_{i}
W_i^{(-)}W_i^{(-)} D_i \varphi^* D_i \varphi
\notag \\
&&+
\cQ_5 (L)
\left( 
\bm{W}^{(-)}
\cdot
\bm{W}^{(-)}
\right)
\left( 
\bm{W}^{(-)}
\cdot
\bm{J}
\right)
+
\cQ_6(L)
\sum_{i}
W_i^{(-)} W_i^{(-)} W_i^{(-)} J_i 
+ \ldots
.\end{eqnarray}
Above we have employed the current operator $\bm{J}$, given by
$\bm{J} = i \left[(\bm{D} \varphi^*) \varphi  -  \varphi^* (\bm{D} \varphi)\right]$. 
A number of things about Eq.~\eqref{eq:fullmess} must be clarified. The 
$\ldots$ 
denotes that we have not finished writing the general Lagrangian allowed by symmetries.
The most general Lagrangian contains a tower of terms with $n$ insertions of 
$W_i^{(-)}$ operators. Writing down all such terms consistent with 
$S_4$ 
for a given 
$n$ 
is arbitrarily complicated.
Fortunately the series expansion of $W_i^{(-)}$ in terms of the gauge field
starts out at a single zero-frequency photon. Thus operators with $n$ insertions
of $W_i^{(-)}$ contribute to processes with at least $n$ zero-frequency 
photons.  In Eq.~\eqref{eq:fullmess}, we have written down all operators 
with at most three insertions of $W_i^{(-)}$. 
Thus the Lagrangian generates all possible couplings to 
\emph{at most} three zero-frequency photons. 
We have also restricted the dynamics to ultra-low energies, so have only kept 
terms with at most two derivatives, $D$, acting on $\varphi$. 
Finally while a term of the form, 
$\left( \bm{W}^{(-)} \cdot \bm{W}^{(-)} \right) |D_\mu \varphi|^2$, 
is allowed by symmetries, it has been removed by a field redefinition.

The coefficients $\cQ_j (L)$ in Eq.~\eqref{eq:fullmess} 
must be determined from 
matching, and thus in general require the calculation of loop graphs 
with an arbitrary number of photons in 
the microscopic theory. 
It is possible that certain additional symmetries
of the underlying theory constrain some coefficients to vanish.
Because this is the zero-frequency sector of an effective theory for a stable particle, 
no multi-particle production thresholds can be attained in loop graphs
that determine the matching coefficients. 
Thus in asymptotically large volumes, the new coupling constants 
$\cQ_j(L)$
will be exponentially small~\cite{Luscher:1990ux,Luscher:1991cf,Lellouch:2000pv}.
Consequently $SO(4)$ will be restored in large volumes.

\subsection{Zero frequency effective theories} \label{s:EFT}

Using the general analysis from above, it is 
straightforward to construct single particle effective
theories that reproduce the zero-frequency results 
derived in Sec.~\ref{s:pions}. 
There is one difference, however.
The underlying theory, QCD, has quark fields
with fractional charges.
Maintaining periodicity of both 
quark fields
under zero-mode gauge transformations requires
a slightly modified quantization condition, namely
\begin{equation} \label{eq:key2}
\bm{\a} = \frac{6 \pi}{e L} \bm{n}
.\end{equation}
This modification reflects that both quark charges 
are quantized in units of $e/3$.
The Wilson lines $\cW_i$ are now given by
\begin{equation}
\cW_i = \exp \left (\frac{i e}{3} \oint d x_i \, A_i \right) 
,\end{equation}
and similarly for the $W_i$. 
Thus for charged and neutral pions,\footnote{%
Because electromagnetism explicitly breaks 
isospin symmetry, we should formulate the low-energy
theories for charged and neutral pions separately. 
Although we utilize
traces over $SU(2)$ 
pion fields $\Phi(\bm{x},t)$,
there are no interactions between pions in Eq.~\eqref{eq:pionEFT}.
}
we require the effective Lagrangian
\begin{eqnarray} \label{eq:pionEFT}
\cL 
&=& 
\frac{1}{2} 
\tr 
(D_\mu \Phi D^\mu \Phi )
-
\frac{1}{2}
m_\pi(L)^2 
\tr (\Phi^2) 
+ i 
\cQ(L) \,
\bm{W}^{(-)}
\cdot 
\tr 
\left[ 
Q (\bm{D} \Phi) \Phi
- 
Q \Phi (\bm{D} \Phi)  
\right]
\notag \\
&&+
\ol \cQ (L)^2
\, 
\bm{W}^{(-)}
\cdot
\bm{W}^{(-)}
\,
\tr 
\left[
(Q \Phi)^2
+ 
\frac{4}{5} 
Q^2 \Phi^2
\right] 
- 
\tilde{\cQ}(L)^2 
\, 
\bm{W}^{(-)}
\cdot
\bm{W}^{(-)}
\,
\tr 
\left[
(Q \Phi)^2
- 
Q^2 \Phi^2
\right] 
,
\notag\\
\end{eqnarray}
where 
$m_\pi(L)$ 
includes the infrared running of the pion mass
(not calculated here), and the new coupling constants 
$\cQ(L)$, 
$\ol \cQ (L)$,
and
$\tilde{\cQ}(L)$ 
are given by
\begin{eqnarray} \label{eq:crazy}
\cQ (L) 
&=& 
-\frac{1}{f^2 L}
\left[ 
2 I_{1/2} (m_\pi^2, L)
+ 
m_\pi^2 
I_{3/2} (m_\pi^2, L)
\right]
\\
\ol \cQ (L)^2 
&=& 
\frac{9}{2 f^2 L^2} m_\pi^4 I_{5/2} (m_\pi^2, L)
\\
\tilde{\cQ}(L) &=& 0
.\end{eqnarray}
For the charged pions, 
we have also calculated all two-photon 
graphs to one-loop order using Eq.~\eqref{eq:L}
(this includes the one-pion irreducible contributions
shown in Fig.~\ref{f:Compton}, and additionally 
the set of one-pion reducible diagrams which are not depicted)
and do not find the need for an extra two-photon 
coupling to charged pions in Eq.~\eqref{eq:pionEFT}.
For this reason, the coupling constant $\tilde{\cQ}(L)$ vanishes. 
The term with coefficient $\ol \cQ(L)$ only 
couples zero-frequency photons to neutral pions.

The single particle effective theory
described by Eq.~\eqref{eq:pionEFT} correctly
reproduces the infrared running of one- and two-photon
processes for both charged and neutral pions at zero frequency. 
This theory is gauge invariant in finite volume
because of the allowance for new operators
which are Wilson lines that cycle the compact dimensions.
These operators, moreover, lead to violation of 
$SO(4)$ 
invariance. Consider a charged pion at rest, 
$\bm{P} = \bm{0}$. 
The current in Eq.~\eqref{eq:chargevertex}
is
\begin{equation} \label{eq:chargevertexrest}
\langle \pi^\pm (\bm{0}) | J_\mu | \pi^\pm (\bm{0}) \rangle 
=  
2 m_\pi \, (\pm e)  \, g_{\mu 0} 
,\end{equation}
where 
$2 m_\pi$ 
is a relativistic normalization factor. 
Boosting to a frame where 
$\bm{P} \neq \bm{0}$, 
generates a current
\begin{equation} \label{eq:currentvertex}
\langle \pi^\pm (\bm{P}) | \bm{J} | \pi^\pm (\bm{P}) \rangle 
=  
2 \sqrt{\bm{P}^2 + m_\pi^2} \, (\pm e)[1 - \cQ(L)]  \, \bm{V}
,\end{equation}
where 
$\bm{V} = \bm{P} / \sqrt{\bm{P}^2 + m_\pi^2}$ 
is the relativistic velocity.  Because of 
$SO(4)$
breaking, the current in this frame 
is not simply the charge times the velocity, 
$\bm{J} \neq (\pm e) \bm{V}$. 
Instead, the current is screened by finite volume effects, 
$\bm{J} = (\pm e) [ 1 - \cQ(L)] \bm{V}$.

\section{Field Theory Identities} \label{s:identity}

Above we have derived finite volume modifications to 
the current of the charged pions and the 
zero-frequency scattering tensor for the neutral 
pions. While we have accounted for these findings using
gauge invariant single particle effective theories, 
here we show that our results are completely consistent
with field theoretic identities valid 
in finite volume.

\subsection{Electromagnetic Vertex}

The zero-frequency part of the 
electromagnetic vertex is constrained by gauge invariance 
via the Ward identity~\cite{Ward:1950xp}. 
Let $\Gamma^\mu (P,P)$ denote the zero-frequency electromagnetic
vertex function of the charged pions. 
The Ward identity requires
\begin{equation} \label{eq:Ward}
-i \Gamma^\mu (P,P) = Q e \, \frac{\partial}{\partial P_\mu} G(P)^{-1}
,\end{equation}
where 
$G(P)$ 
is the pion propagator. In finite volume, we found that the 
wave function correction did
not exactly cancel the forward part of the vertex function. 
This lead to the new coupling
$\cQ(L)$ 
in Eq.~\eqref{eq:crazy}.
Thus at finite volume, the differential form of the Ward identity shown in Eq.~\eqref{eq:Ward} is violated.
Quite simply, however, the steps used to derive Eq.~\eqref{eq:Ward}
are not valid in a fixed finite volume.

On the other hand,
starting from the Ward-Takahashi identity~\cite{Green:1953aa,Takahashi:1957aa} we have 
\begin{equation} \label{eq:Ward2}
-i k_\mu \Gamma^\mu (P+k, P) =  Q e \left[ G(P+k)^{-1}  -  G(P)^{-1} \right]
.\end{equation}
This identity is valid in finite volume. We can demonstrate this explicitly 
using the charged pion vertex function,
$\Gamma^\mu (P+k, P) = \langle \pi^\pm (P+k) | J^\mu | \pi^\pm (P) \rangle$. 
To one-loop order, we evaluate the diagrams in Fig.~\ref{f:Charge}
and contract with the momentum transfer, $k_\mu$. We find 
\begin{eqnarray} 
&& -i k_\mu \langle \pi^\pm (P+k) | J^\mu | \pi^\pm (P) \rangle 
\notag \\
&& \phantom{spac}=  
\mp i e  
\Bigg\{
(2 P + k ) \cdot k  - \frac{2 i}{f^2} \sumint_q 
\left[ \frac{(2 P + k ) \cdot k}{q^2 - m_\pi^2} 
-
\frac{(2 P + k) \cdot q \, (2 q + k ) \cdot k  }{[q^2 - m_\pi^2] [(q+k)^2 - m_\pi^2]}
\right]
\Bigg\} \label{eq:pionWTI}
,\end{eqnarray}
where we have abbreviated
\begin{equation}
\sumint_q \equiv  \frac{1}{L^3} \sum_{\bm{q}= 2 \pi \bm{n} / L} \int \frac{dq^0}{2 \pi}
,\end{equation}
and implicitly regulate ultraviolet divergences using dimensional regularization.
We then write 
\begin{equation} \notag
(2 q + k) \cdot k = (q+k)^2 - m_\pi^2 - [q^2 - m_\pi^2],
\end{equation}
in order to reduce factors in the numerator of the last term. 
To arrive at the Ward-Takahashi identity from Eq.~\eqref{eq:pionWTI}, we must show that
the terms in the $\sumint$\ vanish. This follows immediately by using discrete translational
invariance to re-index the sum. If $\bm{k} \neq 2 \pi \bm{m} / L$, then the summation
over spatial momentum modes cannot be re-indexed in this manner.
Consequently the validity of the Ward-Takahashi identity, Eq.~\eqref{eq:Ward2},
in finite volume hinges on quantized photon momentum.

Having established that the Ward-Takahashi identity holds 
in finite volume, there must be a flaw in the subsequent derivation of the Ward identity.
To arrive at the differential form of the identity, Eq.~\eqref{eq:Ward}, from 
Eq.~\eqref{eq:Ward2} a limiting process $k_\mu \to 0$ is required. 
At fixed volume, the spatial momentum quantization condition invalidates 
this procedure. Contrary to Eq.~\eqref{eq:Ward}, 
there is no condition imposed on $\Gamma^\mu (P,P)$ in a compact space. 
In finite volume with infinite time extent, 
only the spatial part of the differential form of the Ward 
identity does not hold. One can take the limiting procedure
with respect to the zeroth component of momentum transfer, $k^0 \to 0$.
Consequently the time component of Eq.~\eqref{eq:Ward} remains valid.
Our results are indeed consistent with this fact, \emph{cf}. Eq.~\eqref{eq:charged}.\footnote{%
\CPT\ studies of the volume effects for form factors
of pseudoscalar mesons~\cite{Bunton:2006va,Jiang:2006gn} 
have utilized only the time-component of the current,
and considered the extent of the time direction as infinite. 
In this framework, no modification to meson charges was found,
consistent with Eq.~\eqref{eq:Ward}.  
}

\subsection{Compton Tensor}

The classical Thomson cross section arises 
in the zero-frequency limit of 
electromagnetic waves scattering off charged particles. 
According to low-energy theorems~\cite{Gell-Mann:1954kc,Low:1954kd}, 
any sensible gauge invariant field theory of charged particles
will reproduce the Thomson cross section. 
In terms of the off-shell Compton scattering amplitude
for a scalar particle,
the zero-frequency part is required 
to be of the form
\begin{equation} \label{eq:Thomson}
T_{\mu \nu}(\omega = \omega' = 0)
= 
2 (Q e)^2  
\left( g_{\mu \nu}
-
\frac{4  P_\mu P_\nu}{P^2 - m^2}
\right)
,\end{equation}
where $P_\mu$ is the particle's four momentum. 
Upon squaring and multiplying with phase space factors, 
the first term produces the Thomson cross section, 
while the second term is the Born contribution
(which survives when we take the zero frequency limit
\emph{before} going on-shell).
For the neutral pion, Eq.~\eqref{eq:Thomson}
mandates that the Compton tensor vanishes, 
contrary to our results in finite volume, Eq.~\eqref{eq:neutral}.

The Thomson limit of the Compton tensor can be derived
rigorously in field theory from generalized Ward 
identities, specifically for a scalar particle 
we have
\begin{equation} \label{eq:Wardnew}
i T_{\mu \nu}(\omega = \omega' = 0)
= 
(Q e)^2 
G(P)^{-1}
\left[
\frac{\partial^2}{\partial P^\mu \partial P^\nu}
G(P)
\right]
G(P)^{-1}
,\end{equation}
which reproduces both the Thomson and Born terms.
This generalized Ward identity for the two-photon 
amplitude is not valid in finite volume; because,
as with its counterpart in Eq.~\eqref{eq:Ward}, its
derivation relies on a limiting procedure.

Returning to the step in the derivation of 
Eq.~\eqref{eq:Wardnew} before the limiting procedure, we
have a version of the Ward-Takahashi 
identity that is valid in finite volume. 
Let the initial particle (photon) 
momentum be denoted by $P$ ($k$), and 
the final particle (photon) momentum by $P'$ ($k'$). 
Then we have
\begin{equation} \label{eq:WTI}
k'^\nu \, k^\mu \, 
G(P') i T_{\mu \nu} (P',k';k,P) \, G(P)
= 
(Q e)^2 
\Big[
G(P+k) 
-
G(P')
- 
G(P) 
+
G(P - k')
\Big]
.\end{equation}
Using the analytic expression for the one-loop diagrams
in Fig.~\ref{f:Compton} for the neutral pion, 
one can verify explicitly that Eq.~\eqref{eq:WTI}
holds in finite volume provided that 
the photon momenta, $\bm{k}$ and $\bm{k}'$, are quantized.
The validity of Ward-Takahashi identities requires
discrete translational invariance.

Now by taking the limit $k' \to k$, followed by $k \to 0$
in Eq.~\eqref{eq:WTI}, 
we accordingly recover the differential form of the identity 
in Eq.~\eqref{eq:Wardnew}.
Quite simply then, the zero frequency part of the Compton
tensor is not constrained in finite volume as the limit $k \to 0$ cannot be taken. 
Gauge symmetry constrains only the frequency 
dependent combination appearing in Eq.~\eqref{eq:WTI}.
Because we have kept the time direction infinite, 
a limiting procedure does exist for the time-time component 
of the scattering tensor. Consequently Eq.~\eqref{eq:Wardnew}
must apply to $T_{00}(\omega = \omega' =0)$, as is indeed the case 
for our one-loop results for the neutral pion, Eq.~\eqref{eq:neutral}.\footnote{%
By taking the limit $k^0 \to 0$ in 
the singly contracted identity
\begin{equation}
k^\mu G(P') i T_{\mu \nu} G(P)
=
Q e
\left[
G(P') i \Gamma_\nu(P',P+k) G(P+k)
- 
G(P' -k) i \Gamma_\nu(P'-k,P) G(P)
\right]
,\end{equation}
we additionally see that 
$T_{0 \nu}(\omega = \omega' = 0) = 0$
for the neutral pion in finite volume with infinite time extent. 
Similarly the other singly contracted identity
yields 
$T_{\mu 0}(\omega = \omega' = 0) = 0$ upon taking the 
limit $k'^0$ to zero.
Both of these conditions are satisfied by our one-loop results,
Eq.~\eqref{eq:neutral}.
}

\section{Conclusion} \label{s:summy}

Above we have considered infrared effects
on currents in finite volume field theories.
Using the chiral Lagrangian as an example, we showed
that matter fields can additively 
renormalize electric current in finite volume. 
Such effects do not violate gauge invariance; 
on the contrary, new couplings are allowed
because of periodicity constraints on zero-mode
gauge field transformations.
Consequently gauge invariant single particle
effective theories can be formulated that reproduce
the infrared behavior of the interacting theory. 
These theories are written in terms of Wilson lines
that cycle over the compact dimensions.
As $SO(4)$ is explicitly broken in these theories, 
boosting a charged particle from its rest frame to a frame moving with 
velocity $\bm{V}$ does not result in a current $\bm{J} = Q e \bm{V}$.
There are no contradictions with Ward-Takahashi identities,
or low-energy theorems. Differential forms of Ward identities
are inapplicable in finite volume.

Conserved currents are not protected from 
infrared renormalization in finite spaces
with discrete translational invariance. 
As non-perturbative field theories, such as QCD, 
are numerically simulated
in a finite Euclidean space, it is important 
to understand the infrared running of current couplings. 
As a practical application of our work, the single particle
effective theory derived here can be extended 
to describe volume effects for properties of 
hadrons determined from lattice QCD.

\begin{acknowledgments}
We thank W.~Detmold, B.~Smigielski, and especially T.~Mehen for various discussions.
This work is supported in part by the U.S.\ Dept.~of Energy,
Grant No.\ DE-FG02-05ER41368-0 (J.H. and B.C.T.) and by the
Schweizerischer Nationalfonds (F.-J.J.). 
\end{acknowledgments}

\appendix

\section*{Finite Volume Functions}

For processes without momentum insertion, 
all finite volume matching terms, $\D X(L)$, in Eq.~\eqref{eq:cartoon} 
can be cast in terms of the basic building block
\begin{eqnarray}
I_\beta (m^2, L) 
&=& 
\frac{1}{L^3} \sum_{\bm{q}} \frac{1}{[\bm{q}^2 + m^2]^\beta} 
- 
\int \frac{d\bm{q}}{(2 \pi)^3} 
\frac{1}{[\bm{q}^2 + m^2]^\beta} 
\\
&=&
\frac{(m^2)^{3/2 - \b}}{ (4 \pi)^{3/2} \Gamma(\beta)}
\int_0^\infty d\tau
\tau^{\beta - 5/2} e^{- \tau} 
\left[
\vartheta_3(0,e^{- m^2 L^2 / 4 \tau})^3 - 1
\right],
\end{eqnarray}
where $\vartheta_3 (z,q)$ is a Jacobi theta function.
To see that all other required finite volume functions
can be written in terms of $I_\beta(m^2,L)$, we first define
\begin{equation}
I_\beta^{i_1 \ldots i_n} (m^2, L)
=
\frac{1}{L^3} \sum_{\bm{q}} \frac{q^{i_1} \cdots q^{i_n}}
{[\bm{q}^2 + m^2]^\beta} 
- 
\int \frac{d\bm{q}}{(2 \pi)^3} 
\frac{q^{i_1} \cdots q^{i_n}}{[\bm{q}^2 + m^2]^\beta} 
.\end{equation}
As a consequence of cubic invariance in the sums,
we have $I_\beta^{i_1 \ldots i_n} (m^2, L) = 0$, for $n$ odd. 
For even values, we find
\begin{eqnarray}
I_\beta^{i_1 \ldots i_{2n}} (m^2, L)
&=&
\frac{1}{2 n + 1} 
\d^{\{ i_1 i_2 }
\cdots
\d^{i_{2n-1} i_{2n}\}} 
\sum_{j = 0}^{n}
\begin{pmatrix}
n \\
j
\end{pmatrix}
(-m^2)^j I_{\b + j - n} (m^2, L)
.\end{eqnarray}
The bracketed indices denote complete symmetrization
in the usual way, e.g., $\{ A B \} = \frac{1}{2!} (A B + B A)$.

\bibliography{hb}

\end{document}